\documentclass[12pt]{article}

\newcommand{\beq}{\begin{eqnarray}}
\newcommand{\eeq}{\end{eqnarray}}
\newcommand{\beqnn}{\begin{eqnarray*}}
\newcommand{\eeqnn}{\end{eqnarray*}}

\newtheorem{lemma}{Lemma}

\newcommand{\proof}{\paragraph*{{\it Proof.\/}}}

\newcommand{\qed}{\fbox{\phantom{-}}\bigskip}
\newcommand{\rd}{\partial}

\newcommand{\ZZ}{\mathbf{Z}}

\begin{document}


\title{$q$-analogue of modified KP hierarchy \\
and its quasi-classical limit}
\author{Kanehisa Takasaki\\
{\normalsize Graduate School of Human and Environmental Studies, 
Kyoto University}\\
{\normalsize Yoshida, Sakyo, Kyoto 606-8501, Japan}\\
{\normalsize E-mail: takasaki@math.h.kyoto-u.ac.jp}}
\date{}
\maketitle

\begin{abstract}
A $q$-analogue of the tau function of the modified KP hierarchy 
is defined by a change of independent variables.  This tau function 
satisfies a system of bilinear $q$-difference equations.  
These bilinear equations are translated to the language of 
wave functions, which turn out to satisfy a system of linear 
$q$-difference equations.  These linear $q$-difference equations 
are used to formulate the Lax formalism and the description of 
quasi-classical limit.  These results can be generalized to 
a $q$-analogue of the Toda hierarchy.  The results on the $q$-analogue 
of the Toda hierarchy might have an application to the random partition 
calculus in gauge theories and topological strings.  
\end{abstract}
\bigskip

\begin{flushleft}
Running head: $q$-analogue of mKP hierarchy and quasi-classical limit
\end{flushleft}
\bigskip

\begin{flushleft}
arXiv:nlin.SI/0412067
\end{flushleft}
\newpage


\section{Introduction}

The notion of $q$-analogues (also called 
$q$-deformations) of soliton equations has been 
studied from a variety of points of view.  
Kajiwara and Satsuma \cite{bib:KS91} obtained 
a $q$-analogue of the $1 + 1$ dimensional Toda equation 
in a bilinear form.  Partly motivated by this work, 
Mironov, Morozov and Vinet \cite{bib:MMV94} introduced 
a $q$-analogue of the tau function of the Toda hierarchy. 
On the other hand, Wu, Zhang and Zheng \cite{bib:WZZ94} 
found a $q$-analogue of the KdV hierarchy along with 
soliton solutions.  Frenkel \cite{bib:Frenkel96} 
addressed this issue from the point of view of 
$W$-algebras, and considered a $q$-analogue of 
the (generalized) KdV hierarchy and some other 
related soliton equaitons.  Khesin, Lyubashenko 
and Roger \cite{bib:KLR97} proposed a slightly 
different framework of $q$-pseudodifferential 
operators to formulate the Lax representation 
of these $q$-deformed soliton equations. 
Mas and Seco \cite{bib:MS96} applied the work 
of Khesin et al. to a class of $q$-analogues 
of $W$-algebras.  Adler, Horozov and 
van Moerbeke \cite{bib:AHvM98} pointed out that 
these $q$-analogues are closely related to 
the ordinary KP and Toda hierarchies. 
Having obtained a similar result in the context 
of bispectrality \cite{bib:HI97}, Iliev constructed 
a $q$-analogue of the KP hierarchy \cite{bib:Iliev98}.  
Tu \cite{bib:Tu99} studied additional symmetries 
of Iliev's hierarchy.  

In this paper we consider a $q$-analogue of 
the modified KP (mKP for short) hierarchy.  
Our motivation is different from the preceding studies.  
Namely, we are interested in quasi-classical limit 
of the $q$-analogue as the underlying Planck constant 
$\hbar$ tends to $0$.  Actually, our true concern 
lies in the fate of the aforementioned $q$-analogue 
of the Toda hierarchy in the quasi-classical limit.  
The mKP hierarchy, originaly formulated as a system 
of bilinear equations connecting KP tau functions 
\cite{bib:KM81,bib:DJKM81}, may be thought of as 
a subset of the Toda hierarchy \cite{bib:UT84}.  
Because of this, one can use the mKP hierarchy 
as a prototype of technically more complicated 
consideration on the Toda hierarchy.  

In quasi-classical limit, the ordinary Toda hierarchy turns 
into the dispersionless Toda hierarchy \cite{bib:TT93}.  
Accordingly, quasi-classical limit of the mKP hierarchy can be 
realized as a subsystem of the dispersionless Toda hierarchy.  
It is natural to expect a similar result for the $q$-analogue.  
To address this issue, we need a Lax formalism.  We shall 
construct such a Lax formalism in the same framework as 
the ordinary Toda hierarchy.  

A few comments on the mKP hierarchy are in order.  
Firstly, various Lax representations other than ours 
have been proposed for the mKP hierarchy 
\cite{bib:MPZ97,bib:Kupershmidt00,bib:Dickey99}.  
Secondly, the mKP hierarchy also appears in the work of 
Adler et al. \cite{bib:AHvM98} as the ``one-Toda lattice'' 
in their terminology.  Thirdly, Takebe \cite{bib:Takebe02} 
studied a generalization of the mKP hierarchy and 
its dispersionless limit.  His construction is based on 
a different Lax formalism due to Dickey \cite{bib:Dickey99}.  

This paper is organized as follows.  Section 2 is a brief 
review on the tau function of the ordinary mKP hierarchy.  
In Section 3, its $q$-analogue is introduced and shown 
to satisfy a system of bilinear $q$-difference equations.  
In Section 4, these bilinear equations are translated to 
the language of wave functions, which turn out to satisfy 
a system of linear $q$-difference equations.  These linear 
$q$-difference equations are used in Sections 5 and 6 
to formulate the Lax formalism and the description of 
quasi-classical limit.  Section 7 is devoted to the case 
of the $q$-analogue of the Toda hierarchy.

\section{Tau function of modified KP hierarchy} 

The tau function $\tau(s,t)$ of the mKP hierarchy depends 
on a discrete ($\ZZ$ valued) variable $s$ and a set of 
continuous variables $t = (t_1,t_2,\ldots)$, and 
satisfies the bilinear equations 
\beq
  \oint_{\lambda=\infty} \tau(s',t' - [\lambda^{-1}]) 
     \tau(s,t + [\lambda^{-1}])\lambda^{s'-s}e^{\xi(t'-t,\lambda)}
     d\lambda = 0  
\eeq
for $s' \ge s$ and arbitrary values of $t'$ and $t$ 
\cite{bib:KM81,bib:DJKM81}. The contour of the integral 
is understood to be a circle surrounding $\lambda = \infty$, 
and the following standard notations are used:  
\beqnn
  [\alpha] = (\alpha,\alpha^2/2,\ldots,\alpha^k/k,\ldots), \quad 
  \xi(t,\lambda) = \sum_{n=1}^\infty t_n\lambda^n. 
\eeqnn
The bilinear equation for $s' = s$ coincides with that of 
the KP hierarchy, so that $\tau(s,t)$ is a tau function 
of the KP hierarchy.  The bilinear equations for $s' \not= s$ 
define a relation (a B\"acklund transformation 
in a generalized sense) connecting two solutions 
of the KP hierarchy.  A pair of KP tau functions 
in such a relation is generated, for instance, 
by the action of a vertex operator \cite{bib:KM81,bib:DJKM81}. 
If the chain of those KP tau functions $\tau(s,t)$ 
is periodic, i.e., $\tau(s+N,t) = \tau(s,t)$ 
for a positive integer $N$, the relation reduces 
to the ordinary cyclic Darboux transformations 
for the $N$-th generalized KdV hierarchy.  
In the simplest case of $N = 2$, the reduced hierarchy 
with two tau functions  $\tau(0,t)$ and $\tau(1,t)$ 
is nothing but the modified KdV hierarchy.   
This explains the origin of the word ``modified'' 
in the name of the hierarchy.  

Actually, yet another ``modified'' set of 
bilinear equations can be derived from 
these fundamental bilinear equations by 
shifting $t'$ as 
\beqnn
  t' \to t' - [\lambda_1^{-1}] - \cdots - [\lambda_n^{-1}], 
\eeqnn
$\lambda_1,\ldots,\lambda_n$ being assumed to be 
on the far side (i.e., in the domain that contains 
$\lambda = \infty$) of the integration contour.  
The exponential function $e^{\xi(t'-t,\lambda)}$ 
thereby gives rise to an extra factor 
\beqnn
  \prod_{j=1}^n \exp\left(
    - \sum_{k=1}^\infty\frac{\lambda^k}{k\lambda_j^k}\right) 
  = \prod_{j=1}^n \left(1 - \frac{\lambda}{\lambda_j}\right), 
\eeqnn
so that the bilinear equations take 
the ``second-modified'' form 
\beq
  \oint_{\lambda=\infty} 
    \tau(s', t'-[\lambda_1^{-1}]-\cdots-[\lambda_n^{-n}]-[\lambda^{-1}]) 
    \tau(s, t+[\lambda^{-1}]) \nonumber \\
    \mbox{} \times \lambda^{s'-s}e^{\xi(t'-t,\lambda)} 
    (\lambda - \lambda_1)\cdots(\lambda - \lambda_n)d\lambda = 0. 
  \label{eq:mod-bilin}
\eeq
If $\lambda_j$'s are suitably chosen, as we shall see, 
these bilinear equations turn into $q$-difference equations.

\section{$q$-analogue of mKP tau functions}

We now introduce a new set of continuous variables 
$x = (x_1,x_2,\ldots)$ and a set of parameters 
$q = (q_1,q_2,\ldots)$ with $|q_1| < 1$, $|q_2| < 1$, 
etc., and consider the multivariate deformation 
\beq
  \tau_q(s,t,x) 
  = \tau(s,t + \sum_{n=1}^\infty [x_n]_{q_n}^{(n)}) 
\eeq
as a $q$-analogue of the foregoing mKP tau function.  
In this definition, $[\alpha]_q^{(n)}$ stands for 
a $q$-analogue of the $[\alpha]$ symbol of the form 
\beqnn
  [\alpha]_q^{(n)} 
  = \left(0,\ldots,0,\alpha,0,\ldots,0,\frac{(1-q)^2\alpha^2}{2(1-q^2)}, 
    \ldots,0,\ldots,0,\frac{(1-q)^k\alpha^k}{k(1-q^k)},\ldots\right) 
\eeqnn
with the non-zero elements $(1-q)^k\alpha^k/k(1-q^k)$ 
placed in the $kn$-th component for $k = 1,2,\ldots$.  
This is essentially the same change of variables 
as proposed by Mironov et al. \cite{bib:MMV94} 
for the case of the Toda hierarchy, except that 
we keep the old variables $t_1,\,t_2,\,\ldots$ 
along with the new ones $x_1,\,x_2,\,\ldots$.  
If the $x_n$'s other than $x_1$ are set to zero, 
$\tau_q(s,t,x)$ reduces to the univariate deformation 
\beq
  \tau_q(t,x) = \tau(t + [x]_q), \quad [x]_q = [x]_q^{(1)}, 
\eeq
of the KP tau function $\tau(t)$ studied by Adler et al. 
\cite{bib:AHvM98} and Iliev \cite{bib:Iliev98}
in their work on a $q$-analogue of the KP hierarchy.   

$[\alpha]_q^{(n)}$ is defined so as to satisfy 
the $q$-difference relation 
\beq
  [q\alpha]_q^{(n)} = [\alpha]_q^{(n)} - [(1-q)\alpha]^{(n)}, 
\eeq
where $[\alpha]^{(n)}$ denotes a modification of $[\alpha]$ 
of the form 
\beqnn
  [\alpha]^{(n)} 
  = \left(0,\ldots,0,\alpha,0,\ldots,0,\frac{\alpha^2}{2},\ldots,
          0,\ldots,0,\frac{\alpha^k}{k},\ldots\right) 
\eeqnn
in which the $k$-th entry $\alpha^k/k$ of $[\alpha]$ 
is relocated to the $kn$-th component and 
the other components are set to zero.  This is 
a generalization of the $q$-difference relation 
of $[x]_q$ that lies in the heart of the work of 
Adler et al. \cite{bib:AHvM98} and Iliev \cite{bib:Iliev98}.  
As noted by Mironov et al. \cite{bib:MMV94}, 
$[\alpha]^{(n)}$ can be expressed as a linear 
combination of $[\alpha]$ symbols, 
\beqnn
  [\alpha]^{(n)} = \sum_{j=1}^n [e^{2\pi ij/n}\alpha^{1/n}], 
\eeqnn
so that the  $q$-difference relation of $[\alpha]_q^{(n)}$ 
can be rewritten as 
\beq
  [q\alpha]_q^{(n)} = [\alpha]_q^{(n)} 
    - \sum_{j=1}^n [e^{2\pi ij/n}(1-q)^{1/n}\alpha^{1/n}]. 
  \label{eq:[]_q-diff}
\eeq

This $q$-difference relation enables us to relate 
the multiplication $x_n \to q_nx_n$ in $\tau_q(s,t,x)$ 
to the shift $t \to t - [\lambda_1^{-1}] - \cdots - [\lambda_n^{-1}]$ 
in (\ref{eq:mod-bilin}).  If $\lambda_j$'s are indeed 
set to the values 
\beqnn
  \lambda_j = e^{-2\pi ij/n}(1 - q_n)^{-1/n}x_n^{-1/n} 
\eeqnn
that can be read off from (\ref{eq:[]_q-diff}), 
the last polynomial factor in (\ref{eq:mod-bilin}) 
takes the remarkably simple form 
\beqnn
  (\lambda - \lambda_1)\cdots(\lambda - \lambda_n) 
  = \lambda^n - (1 - q_n)^{-1}x_n^{-1}. 
\eeqnn
Thus the modified bilinear equation (\ref{eq:mod-bilin}) 
can be converted to the bilinear equation 
\beq
  \oint_{\lambda=\infty} 
    \tau_q(s',t' - [\lambda^{-1}],\ldots,q_nx_n,\ldots)
    \tau_q(s,t + [\lambda^{-1}],\ldots,x_n,\ldots) 
    \nonumber\\
    \mbox{} \times 
    \lambda^{s'-s}e^{\xi(t'-t,\lambda)} 
    (1- (1-q_n)x_n\lambda^n) d\lambda = 0 
\eeq
for $\tau_q(s,t,x)$.  Moreover, iterating (\ref{eq:[]_q-diff}) 
$k$ times gives higher order $q$-difference relations 
\beq
  [q_n^k x_n]_{q_n}^{(n)} = [x_n]_{q_n}^{(n)} 
    - \sum_{j=1}^n\sum_{\ell=0}^{k-1} 
      [e^{2\pi ij/n} (1-q_n)^{1/n}x_n^{1/n}q_n^{\ell/n}], 
\eeq
from which a similar bilinear equation can be derived.  
Actually, one can consider simultaneous $q$-multiplication 
of $x_n$'s as $x_n \to q_n^{k_n}x_n$ (but $k_n = 0$ except 
for a finite number of $n$'s), which leads to bilinear equations 
of the more general form 
\beq
  \oint_{\lambda=\infty}
    \tau_q(s', t' - [\lambda^{-1}], x')\tau_q(s,t + [\lambda^{-1}],x) 
    \lambda^{s'-s}e^{\xi(t'-t,\lambda)} 
    \nonumber\\
    \mbox{} \times 
    \prod_{n=1}^\infty \prod_{\ell_n=0}^{k_n-1} 
      (1 - (1 - q_n)q_n^{\ell_n}x_n\lambda^n) 
    d\lambda = 0, 
  \label{eq:qtau-bilin}
\eeq
where $x' = (x'_1,x'_2,\ldots)$ and $x = (x_1,x_2,\ldots)$ 
are related by nonnegative powers $q_1^{k_1},q_2^{k_2},\ldots$ 
of $q_1,q_2,\ldots$ as 
\beq
  x'_1 = q_1^{k_1}x_1, \; x'_2 = q_2^{k_2}x_2, \; \ldots.  
\eeq

\section{$q$-analogue of mKP Wave functions}

Since the time variables $t$ of the mKP hierarchy are 
retained in the definition of $\tau_q(x,t,x)$, one can 
define an associated wave function $\Psi_q(s,t,x,\lambda)$ 
and its conjugate $\Psi^*_q(s,t,x,\lambda)$ as wave functions 
of the mKP hierarchy: 
\beq
  \Psi_q(s,t,x,\lambda) 
  &=& \frac{\tau_q(s,t - [\lambda^{-1}],x)}{\tau_q(s,t,x)} 
      \lambda^s e^{\xi(t,\lambda)}e_q(x,\lambda), \nonumber\\
  \Psi_q^*(s,t,x,\lambda) 
  &=& \frac{\tau_q(s,t + [\lambda^{-1}],x)}{\tau_q(s,t,x)} 
      \lambda^{-s} e^{-\xi(t,\lambda)}e_q(x,\lambda)^{-1}. 
\eeq
Note that the exponential part $\lambda^{\pm s}e^{\pm\xi(t,x)}$ 
is multiplied by the extra factor 
\beq
  e_q(x,\lambda) = \prod_{n=1}^\infty e_{q_n}^{x_n\lambda^n}, 
\eeq
where $e_q^z$ denotes the so called $q$-exponential function 
\beqnn
  e_q^z = \exp\left(\sum_{k=1}^\infty \frac{(1-q)^kz^k}{k(1-q^k)}\right)
        = \prod_{k=0}^\infty (1 - (1-q)zq^k)^{-1}. 
\eeqnn
This extra factor originates in the shift of $t$ by the sum of 
$[x_n]_{q_n}^{(n)}$ in the definition of $\tau_q(s,t,x)$, 
and behaves exactly like the ordinary exponential factor 
under the action of $q$-difference operators.  

A prototype of this $q$-exponential factor is 
the $q$-exponential function $e_q^{x\lambda}$ that plays 
a central role in the $q$-analogue of the KP hierarchy 
\cite{bib:AHvM98,bib:Iliev98,bib:Tu99}.  
It satisfies the $q$-difference equation 
\beq
  D_q(x)e_q^{x\lambda} = \lambda e_q^{x\lambda} 
\eeq
for the $q$-difference operator $D_q(x)$ that acts on a function 
of $x$ as 
\beqnn
  D_q(x)f(x) = \frac{f(qx) - f(x)}{qx - x}. 
\eeqnn
Let us also mention that the usual Leibniz rule is modified 
to this operator as 
\beq
  D_q(x)(fg) = D_q(x)f\cdot g + T_q(x)f \cdot D_q(x)g, 
\eeq
where $T_q(x)$ denotes the $q$-multiplication operator 
\beqnn
  T_q(x)f(x) = f(qx). 
\eeqnn

By the same token, as implicitly noted in the work of 
Mironov et al. \cite{bib:MMV94}, the $q$-exponential factor 
$e_q(x,\lambda)$ satisfies the $q$-difference equation 
\beq
  D_{q_n}(x_n)e_q(x,\lambda) = \lambda^n e_q(x,\lambda)
\eeq
or, equivalently, 
\beq
  T_{q_n}(x_n)e_q(x,\lambda) = (1 - (1-q_n)x_n\lambda^n)e_q(x,\lambda). 
\eeq
Iterating the last equation yields 
\beqnn
  T_{q_n}(x_n)^{k_n}e_q(x,\lambda) 
  = \prod_{\ell_n=0}^{k_n-1}(1 - (1-q_n)q_n^{\ell_n}x_n\lambda^n) 
    \cdot e_q(x,\lambda). 
\eeqnn
The prefactor on the right hand side coincides with 
the last polynomial factor in (\ref{eq:qtau-bilin}).  
This means that (\ref{eq:qtau-bilin}) is actually 
a bilinear equation for the wave functions: 
\beq
  \oint_{\lambda=\infty} \Psi_q(s',t',x',\lambda) 
    \Psi_q^*(s,t,x,\lambda) d\lambda = 0. 
  \label{eq:Psi-bilin}
\eeq

We can now follow a standard procedure to derive 
a system of linear $q$-difference equations 
for $\Psi_q(s,t,x,\lambda)$ from the bilinear 
equation (\ref{eq:Psi-bilin}).  A technical clue 
is the following lemma.  

\begin{lemma}
If $\Phi(s,t,x,\lambda)$ is a function (or Laurent series) 
of the form 
\beq
  \Phi(s,t,x,\lambda) 
  = \sum_{n=1}^\infty \phi_n(s,t,x)\lambda^{-n} \cdot 
    \lambda^se^{\xi(t,\lambda)}e_q(x,\lambda) 
\eeq
and satisfies the bilinear equation 
\beq
  \oint_{\lambda=\infty} \Phi(s',t,x,\lambda)
    \Psi_q^*(s,t,x,\lambda) d\lambda = 0 
\eeq
for $s' \ge s$ and arbitrary values of $(t,x)$, 
then $\Phi(s,t,x,\lambda) = 0$.  
\end{lemma}

\proof
The bilinear equation in the statement of the lemma 
implies the equations
\beqnn
  \sum_{n=1}^{s'-s+1} \phi_n(s',t,x)w^*_{s'-s-n+1}(s,t,x) = 0 
\eeqnn
for $\phi_n(s,t,x)$'s and the coefficients of 
the Laurent expansion 
\beqnn
  \Psi_q^*(s,t,x,\lambda) 
  = \Bigl(1 + \sum_{n=1}^\infty w^*_n(s,t,x)\lambda^{-n}\Bigr) 
    \lambda^{-s}e^{-\xi(t,\lambda)}e_q(x,\lambda)^{-1} 
\eeqnn
of $\Psi_q^*(s,t,x,\lambda)$.  Starting with 
the case of $s' = s$ where this equation reduces 
to $\phi_1(s,t,x) = 0$, one can easily show 
by induction that $\phi_n(s,t,x) = 0$ for all $n$.  \qed

To derive a linear $q$-difference equation for 
$\Psi_q(s,t,x,\lambda)$, let us note that 
the action of $D_{q_n}(x_n)$ on $\Psi_q(s,t,x,\lambda)$ 
yields a function of the form $(\lambda^n + \cdots)
\lambda^s e^{\xi(t,\lambda)}e_q(x,\lambda)$, where 
$\lambda + \cdots$ is a Laurent series of $\lambda$ 
that starts from $\lambda^n$ and continues to terms 
of $\lambda^{n-1},\lambda^{n-2},\ldots$.  
One can find a difference operator $C_n$ (i.e., a linear 
combination of the shift operators $e^{m\rd_s}$ that act on 
a function of $s$ as $e^{m\rd_s}f(s) = f(s + m)$) of the form 
\beq
  C_n = e^{n\rd_s} + \sum_{m=1}^n c_{nm}(s,t,x)e^{(n-m)\rd_s} 
\eeq
that absorbs the main part of the Laurent expansion 
of $D_{q_n}(x_n)\Psi_q(s,t,x,\lambda)$ as 
\beqnn
  D_{q_n}(x_n)\Psi_q(s,t,x,\lambda) - C_n\Psi_q(s,t,x,\lambda) 
  = O(\lambda^{-1})\lambda^s e^{\xi(t,\lambda)}e_q(x,\lambda). 
\eeqnn
Let $\Phi(s,t,x,\lambda)$ denote this quantity.  
One can derive, from (\ref{eq:Psi-bilin}), such relations as 
\beqnn
  \oint_{\lambda=\infty} T_{q_n}(x_n)\Psi_q(s,t,x,\lambda) 
    \cdot \Psi^*_q(s,x,t,\lambda) d\lambda = 0 
\eeqnn
and 
\beqnn
  \oint_{\lambda=\infty} e^{m\rd_s}\Psi_q(s,t,x,\lambda) 
    \cdot \Psi^*_q(s,t,x,\lambda) d\lambda = 0 
\eeqnn
for $m \ge 0$.  These relations imply that the assumption 
of the aforementioned lemma holds for $\Phi(s,t,x,\lambda)$, 
which thereby turns out to vanishes identically. 
One can thus confirm that $\Psi_q(s,t,x,\lambda)$ satisfies 
the linear $q$-difference equation 
\beq
  D_{q_n}(x_n)\Psi_q(s,t,x,\lambda) = C_n\Psi_q(s,t,x,\lambda). 
  \label{eq:qmkp-lin}
\eeq

\section{$q$-analogue of Lax, Zakharov-Shabat, and Sato equations} 

Consistency of the linear $q$-difference equations 
(\ref{eq:qmkp-lin}) leads to a system of zero-curvature 
equations for $C_n$'s.  To derive the consistency condition, 
we start from the commutativity relation 
\beqnn
    D_{q_n}(x_n)D_{q_m}(x_m)\Psi_q(s,t,x,\lambda) 
  = D_{q_m}(x_m)D_{q_n}(x_n)\Psi_q(s,t,x,\lambda), 
\eeqnn
and calculate both hand sides using (\ref{eq:qmkp-lin}) 
and the $q$-difference Leibniz rule.  The outcome 
is the equation 
\beqnn
  \Bigl(D_{q_n}(x_n)C_m - D_{q_m}(x_m)C_n 
    + T_{q_n}(x_n)C_m \cdot C_n 
    - T_{q_m}(x_m)C_n \cdot C_m \Bigr)\Psi_q(s,t,x,\lambda) = 0, 
\eeqnn
which implies that the $q$-difference analogue 
\beq
  D_{q_n}(x_n)C_m - D_{q_m}(x_m)C_n 
  + T_{q_n}(x_n)C_m \cdot C_n 
  - T_{q_m}(x_m)C_n \cdot C_m = 0 
  \label{eq:qmkp-zs}
\eeq
of the zero-curvature equation (also called 
``Zakharov-Shabat equations'') is satisfied.  

We now introduce the dressing operator 
(also called ``Sato-Wilson operator'') 
\beq
  W = 1 + \sum_{n=1}^\infty w_n(s,t,x)e^{-n\rd_s}, 
\eeq
where $w_n(s,t,x)$ denote the coefficients of the Laurent 
expansion 
\beqnn
  \Psi_q(s,t,x,\lambda) 
  = \Bigl(1 + \sum_{n=1}^\infty w_n(s,t,x)\lambda^{-n}\Bigr) 
    \lambda^s e^{\xi(t,\lambda)}e_q(x,\lambda).  
\eeqnn
(\ref{eq:qmkp-lin}) can be converted to the $q$-difference 
equations 
\beq
  D_{q_n}(x_n)W = C_n W - T_{q_n}(x_n)W \cdot e^{n\rd_s}. 
\eeq
One can eliminate $C_n$ to derive a nonlinear system 
of equations for $W$.  This is achieved by multiplying 
both hand sides by $W$ from the right side, 
\beqnn
  D_{q_n}(x_n)W\cdot W^{-1} 
  = C_n - T_{q_n}(x_n)W \cdot e^{n\rd_s}W^{-1}, 
\eeqnn
and separating the difference operators into 
the part of nonnegative/negative powers of $e^{\rd_s}$.  
Let $(A)_{\ge 0}$ and $(A)_{<0}$ denote the projection 
of a difference operator to these parts: 
\beqnn
  \Bigl(\sum_n a_n e^{n\rd_s}\Bigr)_{\ge 0} 
    = \sum_{n\ge 0}a_n e^{n\rd_s}, \quad 
  \Bigl(\sum_n a_n e^{n\rd_s}\Bigr)_{<0} 
    = \sum_{n<0} a_n e^{n\rd_s}. 
\eeqnn
The $(\cdot)_{\ge 0}$ part of the foregoing $q$-difference 
equation gives the relation 
\beq
  C_n = \Bigl(T_{q_n}(x_n)W \cdot e^{n\rd_s}W^{-1}\Bigr)_{\ge 0}. 
  \label{eq:Cn-W}
\eeq
This enables one to eliminate $C_n$ from the $q$-difference 
equation itself.  The outcome is the nonlinear 
$q$-difference equation 
\beq
  D_{q_n}(x_n)W 
  = - \Bigl(T_{q_n}(x_n)W \cdot e^{n\rd_s}W^{-1}\Bigr)_{<0}W. 
  \label{eq:qmkp-sato}
\eeq
This is a $q$-difference analogue of the so called 
``Sato equations''. 

Finally we introduce the Lax operator 
\beq
  L = W e^{\rd_s}W^{-1} 
    = e^{\rd_s} + \sum_{n=1}^\infty u_n e^{(1-n)\rd_s}, 
\eeq
which turns out to satisfy the $q$-difference 
Lax equations 
\beq
  D_{q_n}(x_n)L = C_n L - T_{q_n}(x_n)L \cdot C_n 
  \label{eq:qmkp-lax}
\eeq
as a consequence of the Sato equations (\ref{eq:qmkp-sato}).  
The same Lax equations can also be derived from 
the consistency of the formal spectral equation 
\beq
  L\Psi_q(s,t,x,\lambda) = \lambda\Psi_q(s,t,x,\lambda) 
\eeq
with the linear $q$-difference equations (\ref{eq:qmkp-lin}).  
The commutativity, $D_{q_n}(x_n)D_{q_m}(x_m)L = D_{q_m}(x_m)L$, 
of the flows of the $q$-difference Lax equations is 
ensured by the $q$-difference Zakharov-Shabat equations 
(\ref{eq:qmkp-zs}).  

To conclude, let us mention that $\Psi_q(s,t,x,\lambda)$ 
satisfies the linear equations 
\beq
  \rd_{t_n}\Psi_q(s,t,x,\lambda) = B_n \Psi_q(s,t,x,\lambda) 
\eeq
of the ordinary mKP hierarchy as well, $B_n$ being 
a difference operator of the form 
\beq
  B_n = e^{n\rd_s} + \sum_{m=1}^n b_{nm}(s,t,x)e^{(n-m)\rd_s}.  
\eeq
One can derive from these linear equations 
the Sato equations
\beq
  \rd_{t_n}W = - \Bigl(W e^{n\rd_s}W^{-1}\Bigr)_{<0}W, 
\eeq
the Zakharov-Shabat equations 
\beq
  \rd_{t_n}B_m - \rd_{t_m}B_n + [B_m,B_n] = 0, 
\eeq
and the Lax equations 
\beq
  \rd_{t_n}L = [B_n,L]
  \label{eq:mkp-lax}
\eeq
along with the relation 
\beq
  B_n = \Bigl(W e^{n\rd_s}W^{-1}\Bigr)_{\ge 0} = (L^n)_{\ge 0} 
  \label{eq:Bn-W}
\eeq
connecting $B_n$ with $W$ and $L$.  These equations 
comprise a Lax formalism of the mKP hierarchy.  
What we have derived above is a $q$-difference analogue 
thereof.

\section{Quasi-classical limit and Hamilton-Jacobi equations}

We now turn to the issue of quasi-classical limit.  
To this end, we set the parameters $q_n$ to 
depend on the Planck constant $\hbar$ as 
\beq
  q_n = q^n = e^{-n\beta\hbar},\quad 
  q = e^{-\beta\hbar},  
\eeq
where $\beta$ is an arbitrary constant, and 
rescale the independent variables $s,t_n,x_n$ as 
\beq
  s \to s/\hbar, \quad 
  t_n \to t_n/\hbar, \quad 
  x_n \to x_n/(1 - q^n). 
\eeq
The fundamental operators $e^{m\rd_s}$, $\rd_{t_n}$ 
and $D_{q^n}(x_n)$  are thereby rescaled as 
\beqnn
  e^{m\rd_s} \to e^{m\hbar\rd_s}, \quad 
  \rd_{t_n} \to \hbar\rd_{t_n}, \quad 
  D_{q^n}(x_n) \to (1 - q^n)D_{q^n}(x_n), 
\eeqnn
which resembles the usual set-up of quantum mechanics.  
The linear equations of the wave functions are 
reformulated in the $\hbar$-dependent form as 
\beq
  \hbar\rd_{t_n}\Psi_q(\lambda) &=& B_n\Psi_q(\lambda)
  \label{eq:hbar-mkp-lin} 
\eeq
and 
\beq
  (1 - q^n)D_{q^n}(x_n)\Psi_q(\lambda) &=& C_n\Psi_q(\lambda).  
  \label{eq:hbar-qmkp-lin}
\eeq 
Also note that we have slightly changed the notation, 
namely, we omit writing the dependence on $x,t,x,\hbar$ 
to highlight the dependence on $\lambda$.  
The coefficients $b_{nm}$ and $c_{nm}$ of 
\beqnn
  B_n = e^{n\hbar\rd_s} + \sum_{m=1}^n b_{nm}e^{(n-m)\hbar\rd_s}, 
  \quad 
  C_n = e^{n\hbar\rd_s} + \sum_{m=1}^n c_{nm}e^{(n-m)\hbar\rd_s}, 
\eeqnn
too, are understood to depend on $\hbar$ arbitrarily. 
To achieve reasonable quasi-classical limit, however, 
we assume that $b_{nm}$ and $c_{nm}$ have a smooth limit 
\beq
  \lim_{\hbar\to 0}b_{nm} = b_{nm}^0, \quad 
  \lim_{\hbar\to 0}c_{nm} = c_{nm}^0 
\eeq
as $\hbar \to 0$.  As in the case of the Toda hierarchy 
\cite{bib:TT93}, this imposes a rather strong condition 
on the solution in consideration;  let us simply assume 
that these conditions are satisfied.  One can readily 
see from (\ref{eq:Cn-W}) and (\ref{eq:Bn-W}) that 
$B_n$ and $C_n$ coincide in the limit as $\hbar \to 0$, 
namely, 
\beq
  b^0_{nm} = c^0_{nm}. 
\eeq

Under these assumptions, one can set the wave function 
in the WKB form 
\beq
  \Psi_q(\lambda) = \exp\Bigl(\hbar^{-1}S(\lambda) + O(\hbar^0)\Bigr) 
\eeq
or, more loosely, 
\beqnn
  \Psi_q(\lambda) \sim e^{\hbar^{-1}S(\lambda)}. 
\eeqnn
Since the $q$-exponential factor $e_q(x,\lambda)$ behaves as 
\beq
  e_q(x,\lambda) = \exp\left(\sum_{n=1}^\infty 
      \frac{\mathrm{Li}_2(x_n\lambda^n)}{n\beta\hbar} + O(\hbar^0) \right), 
\eeq
where $\mathrm{Li}_2(z)$ denotes the dilogarithmic function 
\beqnn
  \mathrm{Li}_2(z) = \sum_{k=1}^\infty \frac{z^k}{k^2}, 
\eeqnn
the phase function $S(\lambda) = S(s,t,x,\lambda)$ is 
a Laurent series of the form 
\beq
  S(\lambda) = s\log\lambda + \xi(t,\lambda) 
    + \sum_{n=1}^\infty \frac{\mathrm{Li}_2(x_n\lambda^n)}{n\beta} 
    + (\mbox{negative powers of $\lambda$}).  
\eeq
One can derive from (\ref{eq:hbar-mkp-lin}) and (\ref{eq:hbar-qmkp-lin}) 
a system of Hamilton-Jacobi equations as follows.  

For warm-up, let us first consider (\ref{eq:hbar-mkp-lin}).  
Actually, this has been done in the study of quasi-classical limit 
of the Toda hierarchy \cite{bib:TT93}.  Upon substituting 
the WKB ansatz, both hand sides of (\ref{eq:hbar-mkp-lin}) 
can be expressed as 
\beqnn
  \rd_{t_n}\Psi_q(\lambda) 
    \sim \rd_{t_n}S(\lambda)e^{\hbar^{-1}S(\lambda)}, \quad 
  B_n\Psi_q(\lambda) 
    \sim \mathcal{B}_n(e^{\rd_sS(\lambda)})e^{\hbar^{-1}S(\lambda)}, 
\eeqnn
where $\mathcal{B}_n(p)$ is a polynomial of the form 
\beq
  \mathcal{B}_n(p) = p^n + \sum_{m=1}^n b_{nm}^0 p^{n-m}. 
\eeq
Thus one obtains the Hamilton-Jacobi equation 
\beq
  \rd_{t_n}S(\lambda) = \mathcal{B}_n(e^{\rd_sS(\lambda)}). 
  \label{eq:mkp-hj}
\eeq

Let us now consider (\ref{eq:hbar-qmkp-lin}). 
The main task is to determine the asymptotic form of 
the left hand side, i.e., 
\beqnn
\lefteqn{
  (1 - q^n)D_{x_n}(q^n)\Psi_s(\lambda) }\nonumber\\
  &=& x_n^{-1}\Bigl(\exp(\hbar^{-1}S(\lambda) + O(\hbar^0)) 
            - \exp(\hbar^{-1}T_{x_n}(q^n)S(\lambda) + O(\hbar^0))\Bigr). 
\eeqnn
$T_{x_n}(q^n)S(\lambda)$ can be expanded to a Taylor series 
of $\hbar$ as 
\beqnn
  T_{x_n}(q^n)S(s,t,x,\lambda) 
  &=& S(s,t,\ldots,e^{-n\beta\hbar}x_n,\ldots,\lambda) \nonumber \\
  &=& S(s,t,x,\lambda) 
      - n\beta\hbar x_n\rd_{x_n}S(s,t,x,\lambda) 
      + O(\hbar^2). 
\eeqnn
Consequently, 
\beqnn
  (1 - q^n)D_{x_n}(q^n)\Psi_s(\lambda) 
  \sim x_n^{-1}\Bigl(1 - \exp(-n\beta x_n\rd_{x_n}S(\lambda))\Bigr) 
       e^{\hbar^{-1}S(\lambda)}. 
\eeqnn
On the other hand, since $c_{nm}$ and $b_{nm}$ coincide 
in the limit as $\hbar \to 0$, 
\beqnn
  C_n\Psi_q(\lambda) 
  \sim \mathcal{B}_n(e^{\rd_sS(\lambda)})e^{\hbar^{-1}S(\lambda)}. 
\eeqnn
Thus one obtains the Hamilton-Jacobi equation 
\beq 
  x_n^{-1}\Bigl(1 - \exp(-n\beta x_n\rd_{x_n}S(\lambda))\Bigr) 
  = \mathcal{B}_n(e^{\rd_sS(\lambda)}). 
\eeq
It will be also suggestive to rewrite this equation as 
\beq
  x_n\rd_{x_n}S(\lambda) = - \frac{1}{n\beta}
    \log(1 - x_n\mathcal{B}_n(e^{\rd_sS(\lambda)})). 
  \label{eq:qmkp-hj}
\eeq

One can now follow the case of the Toda hierarchy \cite{bib:TT93} 
to convert these Hamilton-Jacobi equations to a Lax formalism 
of the ``dispersionless'' type \cite{bib:TT91}.  
The Lax function $\mathcal{L}(p)$  is obtained by 
solving the equation 
\beq
  \rd_sS(\lambda) = p 
\eeq
for $\lambda$ as 
\beq
  \lambda = \mathcal{L}(p) 
  = p + \sum_{n=1}^\infty u_n^0 p^{1-n}. 
\eeq
The coefficients $u_n^0$ coincide with the limit 
of the coefficients $u_n$ of $L$ as $\hbar \to 0$.  
(\ref{eq:mkp-hj}) and (\ref{eq:qmkp-hj}) correspond 
to the Lax equations 
\beq
  \rd_{t_n}\mathcal{L}(p) 
    &=& \{\mathcal{B}_n(p),\mathcal{L}(p)\}, \nonumber\\
  x_n\rd_{x_n}\mathcal{L}(p) 
    &=& - \frac{1}{n\beta}\{\log(1 - x_n\mathcal{B}_n(p)),\mathcal{L}(p)\}, 
\eeq
with respect to the Poisson bracket 
\beqnn
  \{f,g\} = p\rd_p f \cdot \rd_s g - \rd_s f \cdot p\rd_p g. 
\eeqnn
One can further introduce the Orlov-Schulman function 
\beq
  \mathcal{M}(p) 
  = \lambda\rd_\lambda S(\lambda)|_{\lambda=\mathcal{L}(p)}, 
\eeq
which turns out to satisfy the Lax equations 
\beq
  \rd_{t_n}\mathcal{M}(p) 
    &=& \{\mathcal{B}_n(p),\mathcal{M}(p)\}, \nonumber\\ 
  x_n\rd_{x_n}\mathcal{M}(p) 
    &=& - \frac{1}{n\beta}\{\log(1 - x_n\mathcal{B}_n(p)),\mathcal{M}(p)\} 
\eeq
and the Poisson commutation relation 
\beq
  \{\mathcal{L}(p),\mathcal{M}(p)\} = \mathcal{L}(p). 
\eeq

\section{$q$-analogue of Toda hierarchy}

The tau function $\tau(s,t,\bar{t})$ of the Toda hierarchy 
\cite{bib:UT84} depends on yet another set of 
continuous variables $\bar{t} = (\bar{t}_1,\bar{t}_2,\ldots)$, 
and satisfy the bilinear equations 
\beq
\lefteqn{
  \oint_{\lambda=\infty} \tau(s',t'-[\lambda^{-1}],\bar{t}') 
    \tau(s,t+[\lambda^{-1}],\bar{t})
    \lambda^{s'-s}e^{\xi(t'-t,\lambda)}d\lambda }\nonumber \\
  &=& \oint_{\lambda=0} \tau(s',t',\bar{t}'-[\lambda]) 
        \tau(s,t,\bar{t}+[\lambda]) 
        \lambda^{s'-s}e^{\xi(\bar{t}'-\bar{t},\lambda^{-1})}d\lambda, 
\eeq
where $s'$ and $s$ ar now aribrary (namely, there is 
no ordering constraint), and the contours of 
the integrals on both hand sides are understood 
to be a circle surrounding $\lambda = \infty$ and 
$\lambda = 0$ respectively.  Following Mironov, 
Morozov and Vinet \cite{bib:MMV94}, we now introduce 
two sets of continuous variables $x = (x_1,x_2,\ldots)$, 
$\bar{x} = (\bar{x}_1,\bar{x}_2,\ldots)$ along with 
parameters $q = (q_1,q_2,\ldots)$, 
$\bar{q} = (\bar{q}_1,\bar{q}_2,\ldots)$, 
and consider the $q$-analogue 
\beq
  \tau_{q,\bar{q}}(s,t,\bar{t},x,\bar{x}) 
  = \tau(s,\, t + \sum_{n=1}^\infty [x_n]_{q_n}^{(n)},\, 
     \bar{t} + \sum_{n=1}^\infty [\bar{x}_n]_{\bar{q}_n}^{(n)}) 
\eeq
of $\tau(s,t,\bar{t})$.  The modified bilinear equations 
(\ref{eq:mod-bilin}) can be generalized to 
$\tau_{q,\bar{q}}(s,t,\bar{t},x,\bar{x})$.  One can 
rewrite those modified bilinear equations for 
$\tau_{q,\bar{q}}(s,t,\bar{t},x,\bar{x})$ 
to the bilinear equations 
\beq
\lefteqn{
  \oint_{\lambda=\infty}
    \Psi_{q,\bar{q}}(s',t',\bar{t}',x',\bar{x}',\lambda) 
    \Psi_{q,\bar{q}}^*(s,t,\bar{t},x,\bar{x},\lambda)d\lambda }\nonumber\\
  &=& \oint_{\lambda=0}
      \bar{\Psi}_{q,\bar{q}}(s',t',\bar{t}',x',\bar{x}',\lambda) 
      \bar{\Psi}_{q,\bar{q}}^*(s,t,\bar{t},x,\bar{x},\lambda)d\lambda 
\eeq
for the wave functions 
\beq
  \Psi_{q,\bar{q}}(s,t,\bar{t},x,\bar{x},\lambda) 
  &=& \frac{\tau_{q,\bar{q}}(s,t-[\lambda^{-1}],\bar{t},x,\bar{x})}
           {\tau_{q,\bar{q}}(s,t,\bar{t},x,\bar{x})} 
      \lambda^se^{\xi(t,\lambda)}e_q(x,\lambda), \nonumber\\
  \Psi_{q,\bar{q}}^*(s,t,\bar{t},x,\bar{x},\lambda) 
  &=& \frac{\tau_{q,\bar{q}}(s,t+[\lambda^{-1}],\bar{t},x,\bar{x})} 
           {\tau_{q,\bar{q}}(s,t,\bar{t},x,\bar{x})} 
      \lambda^{-s}e^{-\xi(t,\lambda)}e_q(x,\lambda)^{-1}, \nonumber\\
  \bar{\Psi}_{q,\bar{q}}(s,t,\bar{t},x,\bar{x},\lambda) 
  &=& \frac{\tau_{q,\bar{q}}(s+1,t,\bar{t}-[\lambda],x,\bar{x})} 
           {\tau_{q,\bar{q}}(s,t,\bar{t},x,\bar{x})} 
      \lambda^se^{\xi(\bar{t},\lambda^{-1})}
      e_{\bar{q}}(\bar{x},\lambda^{-1}), \nonumber\\
  \bar{\Psi}_{q,\bar{q}}^*(s,t,\bar{t},x,\bar{x},\lambda) 
  &=& \frac{\tau_{q,\bar{q}}(s-1,t,\bar{t}+[\lambda],x,\bar{x})} 
           {\tau_{q,\bar{q}}(s,t,\bar{t},x,\bar{x})} 
      \lambda^{-s}e^{-\xi(\bar{t},\lambda^{-1})}
      e_{\bar{q}}(\bar{x},\lambda^{-1})^{-1}. 
\eeq
$x'_n,\bar{x}'_n,x_n,\bar{x}_n$ are understood 
to be related by nonnegative powers $q_n^{k_n}$ and 
$\bar{q}_n^{\bar{k}_n}$ of $q_n$ and $\bar{q}_n$ as 
\beq
  x'_1 = q_1^{k_1}x_1,\; 
  x'_2 = q_1^{k_2}x_2,\; \ldots, \quad 
  \bar{x}'_1 = \bar{q}_1^{\bar{k}_1}\bar{x}_1,\; 
  \bar{x}'_2 = \bar{q}_2^{\bar{k}_2}\bar{x}_2,\; \ldots. 
\eeq

The wave functions turn out to satisfy a set of 
linear differential and $q$-difference equations. 
To this end, the previous lemma has to be generalized 
to an equation of the form 
\beq
\lefteqn{
  \oint_{\lambda=\infty}\Phi(s',t,\bar{t},x,\bar{x},\lambda) 
    \Psi_{q,\bar{q}}(s,t,\bar{t},x,\bar{x},\lambda)d\lambda }\nonumber\\
  &=& \oint_{\lambda=0}\bar{\Phi}(s',t,\bar{t},x,\bar{x},\lambda) 
    \bar{\Psi}_{q,\bar{q}}(s,t,\bar{t},x,\bar{x},\lambda)d\lambda, 
\eeq
where $\Phi(s,t,\bar{t},x,\bar{x},\lambda)$ 
is the same as in the previous case and 
$\bar{\Phi}(s,t,\bar{t},x,\bar{x},\lambda)$ 
is a function of the form 
\beq
  \bar{\Phi}(s,t,\bar{t},x,\bar{x},\lambda) 
  = \sum_{n=1}^\infty 
    \bar{\phi}_n(s,t,\bar{t},x,\bar{x})\lambda^n \cdot 
    \lambda^se^{\xi(\bar{t},\lambda^{-1})} 
    e_{\bar{q}}(\bar{x},\lambda^{-1}). 
\eeq
The generalized statement is that such a pair of 
functions $\Phi(s,t,\bar{t},x,\bar{t},\lambda)$ 
and $\bar{\Phi}(s,t,\bar{t},x,\bar{t},\lambda)$ 
vanish identically.  One can threby show that 
$\Psi_{q,\bar{q}}(s,t,\bar{t},x,\bar{x},\lambda)$ 
satisfy the linear $q$-difference equations 
\beq
  D_{q_n}(x_n)\Psi_{q,\bar{q}}(s,t,\bar{t},x,\bar{x},\lambda) 
  &=& C_n\Psi_{q,\bar{q}}(s,t,\bar{t},x,\bar{x},\lambda), 
  \nonumber\\
  D_{\bar{q}_n}(\bar{x}_n)\Psi_{q,\bar{q}}(s,t,\bar{t},x,\bar{x},\lambda) 
  &=& \bar{C}_n\Psi_{q,\bar{q}}(s,t,\bar{t},x,\bar{x},\lambda) 
\eeq
along with the linear differential equations 
\beq
  \rd_{t_n}\Psi_{q,\bar{q}}(s,t,\bar{t},x,\bar{x},\lambda) 
  &=& B_n\Psi_{q,\bar{q}}(s,t,\bar{t},x,\bar{x},\lambda), 
  \nonumber\\
  \rd_{\bar{t}_n}\Psi_{q,\bar{q}}(s,t,\bar{t},x,\bar{x},\lambda) 
  &=& \bar{B}_n\Psi_{q,\bar{q}}(s,t,\bar{t},x,\bar{x},\lambda) 
\eeq
of the ordinary Toda hierarchy, and that the same equations hold 
for $\bar{\Psi}_{q,\bar{q}}(s,t,\bar{t},x,\bar{x},\lambda)$ 
as well.  $B_n$ and $C_n$ are the same difference operators 
as in the previously case.  $\bar{B}_n$ and $\bar{C}_n$ 
are difference operators of the form 
\beq
  \bar{B}_n = \sum_{m=0}^{n-1} \bar{b}_{nm}e^{(m-n)\rd_s}, \quad 
  \bar{C}_n = \sum_{m=0}^{n-1} \bar{c}_{nm}e^{(m-n)\rd_s}. 
\eeq

Quasi-classical limit is achieved by setting $q_n = \bar{q}_n 
= q^n$, $q = e^{-\beta\hbar}$, and rescaling the independent 
variables as $s \to s/\hbar$, $t_n \to t_n/\hbar$, 
$\bar{t}_n \to \bar{t}_n/\hbar$, $x_n \to x_n/(1 - q^n)$, 
$\bar{x}_n \to \bar{x}_n/(1 - q^n)$.  Also the coefficients 
of $B_n,\bar{B}_n,C_n,\bar{C}_n$ are allowed to depend on 
$\hbar$ arbitrarily, except that they have a smooth limit 
as $\hbar \to 0$.  Assuming the WKB ansatz 
\beq
  \Psi_{qq}(s,\lambda) 
  \sim \exp\Bigl(\hbar^{-1}S(\lambda) + O(\hbar^0)\Bigr), 
\eeq
one can derive the Hamilton-Jacobi equations 
\beq
  \rd_{t_n}S(\lambda) 
    &=& \mathcal{B}_n(e^{\rd_sS(\lambda)}), \nonumber\\
  \rd_{\bar{t}_n}S(\lambda) 
    &=& \bar{\mathcal{B}}_n(e^{\rd_sS(\lambda)}), \nonumber\\
  x_n\rd_{x_n}S(\lambda) 
    &=& - \frac{1}{n\beta}
          \log(1 - \mathcal{B}_n(e^{\rd_sS(\lambda)})), \nonumber\\
  \bar{x}_n\rd_{\bar{x}_n}S(\lambda) 
    &=& - \frac{1}{n\beta}
          \log(1 - \bar{\mathcal{B}}_n(e^{\rd_sS(\lambda)})), 
\eeq
where $\mathcal{B}_n(p)$ is the same polynomial 
as in the case of the modified KP hierarchy and 
$\bar{\mathcal{B}}_n(p)$ is a polynomial in $p^{-1}$ 
of the form 
\beq
  \bar{\mathcal{B}}_n(p) = \sum_{m=0}^{n-1} \bar{b}_{nm}^0 p^{m-n}, 
  \quad \bar{b}_{nm}^0 = \lim_{\hbar\to 0}\bar{b}_{nm}.  
\eeq
One can define the Lax function 
$\mathcal{L}(p)$ and the Orlov-Schulman function 
$\mathcal{M}(p)$ in the same way as the previous case.  
They satisfy the Poisson commutation relation 
\beq
  \{\mathcal{L}(p),\mathcal{M}(p)\} = \mathcal{L}(p) 
\eeq
and a system of Lax equations with respect to 
the same Poisson bracket $\{\cdot,\cdot\}$.  
The Lax equations for $\mathcal{L}(p)$ read 
\beq
  \rd_{t_n}\mathcal{L}(p) 
   &=& \{\mathcal{B}_n(p),\mathcal{L}(p)\}, \nonumber\\
  \rd_{\bar{t}_n}\mathcal{L}(p) 
   &=& \{\bar{\mathcal{B}}_n(p),\mathcal{L}(p)\}, \nonumber\\
  x_n\rd_{x_n}\mathcal{L}(p) 
   &=& - \frac{1}{n\beta}
         \{\log(1 - x_n\mathcal{B}_n(p)),\mathcal{L}(p)\}, \nonumber\\
  \bar{x}_n\rd_{\bar{x}_n}\mathcal{L}(p) 
   &=& - \frac{1}{n\beta}
         \{\log(1 - \bar{x}_n\bar{\mathcal{B}}_n(p)),\mathcal{L}(p)\}, 
\eeq
and Lax equations of the same form hold for 
$\mathcal{M}(p)$.  One can repeat the same calculations 
for the WKB ansatz 
\beq
  \bar{\Psi}_{q,\bar{q}}(\lambda) 
  \sim \exp\Bigl(\hbar^{-1}\bar{S}(\lambda)  + O(\hbar^0)\Bigr) 
\eeq
of $\bar{\Psi}_{q,\bar{q}}(\lambda)$.  This eventually leads 
to the second pair $\bar{\mathcal{L}}(p),\bar{\mathcal{M}}(p)$ 
of Lax and Orlov-Schulman functions.   Using the four 
functions $\mathcal{L}(p),\mathcal{M}(p),
\bar{\mathcal{L}}(p),\bar{\mathcal{M}}(p)$, 
one can formulate a ``twistor theory'' \cite{bib:TT91} 
of the extended Lax formalism.

\section{Conclusion}

We have considered a $q$-analogue of 
the mKP hierarchy as a prototype of the $q$-analogue 
of the Toda hierarchy.  As expected, we have been able 
to describe the quasi-classical limit of both systems 
by the method developed for the Toda hierarchy \cite{bib:TT93}.  
An unusual feature is that the $q$-difference flows 
turn into the flows with logarithmic generators 
such as $\log(1 - x_n\mathcal{B}_n(p))$ and 
$\log(1 - \bar{x}_n\bar{\mathcal{B}}_n(p))$.  
This is reminiscent of a logarithmic structure 
that can be seen in Takebe's result \cite{bib:Takebe02}.  

As a final remark, let us mention a possible link of 
the present work with the random partition calculus in 
gauge theories and topological strings \cite{bib:NO03,bib:NRV03}.  
Such a link can be seen in the simple (almost trivial) tau function 
\beq
  \tau(s,t,\bar{t}) = \Lambda^{s^2} 
    \exp\left(- \sum_{n=1}^\infty \Lambda^{2n}nt_n\bar{t}_n\right) 
\eeq
of the Toda hierarchy, where $\Lambda$ is an arbitrary constant.  
If $q_n,\bar{q}_n$ are set to the special value $q^n$, and 
all $t_n,\bar{t}_n,x_n,\bar{x}_n$ but $x_1$ and $\bar{x_1}$ 
are turned off, then the $q$-deformed tau function reduces to 
\beq
  \tau_{q,\bar{q}}(s,x_1,\bar{x}_1) 
  = \Lambda^{s^2} \exp\left(- \sum_{k=1}^\infty 
    \frac{(\Lambda^2(1-q)^2x_1\bar{x}_1)^k}{k(1-q^k)^2}\right), 
\eeq
which, upon suitably adjusting the variables $x_1,\bar{x}_1$, 
coincides with a relevant physical quantity (such as 
the partition function) in the random partition calculus.

\subsection*{Acknowledgements}

I would like to thank Toshio Nakatsu and Takashi Takebe 
for discussion and comments.  This work is a byproduct 
of joint research with Nakatsu on random partitions.  
This work was partly supported by the Grant-in-Aid 
for Scientific Research (No. 16340040) from the Ministry 
of Education, Culture, Sports and Technology.


\end{document}